\documentclass[12pt]{iopart}
\usepackage{iopams}  

\usepackage{xcolor}
\usepackage{graphicx}
\usepackage{caption}
\usepackage{subcaption}
\expandafter\let\csname equation*\endcsname\relax

\expandafter\let\csname endequation*\endcsname\relax

\usepackage{amsmath}
\usepackage{mathrsfs}

\usepackage[T1]{fontenc}
\newcommand{\thorn}{\mbox{\th}}
\usepackage{float}
\usepackage{hyperref}
\usepackage[%backend=bibtex,
            style=phys,
            sorting=nyt,
            articletitle=true,
            biblabel=brackets,
            chaptertitle=true,
            pageranges=true,
            hyperref=true,
            maxcitenames=3,
            maxbibnames=4,
            url=true,
            autocite=superscript
            ] {biblatex}

\newcommand*{\hdal}{{\Box\kern-7.5pt\wedge}}

\newcommand*{\scri}{\mathscr{I}}
\newcommand*{\const}{\mathit{const.}}
\newcommand*{\ii}{\mathrm{i}}

\numberwithin{equation}{section}

\begin{document}
\title[The non-linear perturbation of a black hole by gravitational waves]{The non-linear perturbation of a black hole by gravitational waves. III. Newman-Penrose constants}

\author{J Frauendiener$^1$, A Goodenbour$^2$ and C Stevens$^2$}

\address{${}^1$Department of Mathematics and Statistics, University of Otago,
  Dunedin 9016, New Zealand \\ ${}^2$Department of Mathematics and Statistics,
  University of Canterbury, Christchurch 8041, New Zealand}

\ead{joerg.frauendiener@otago.ac.nz, alex.goodenbour@pg.canterbury.ac.nz,
  chris.stevens@canterbury.ac.nz}

\begin{abstract}
In this paper we continue our study of the non-linear response of a Schwarzschild black hole to an ingoing gravitational wave by computing the Newman-Penrose (NP) constants. The NP constants are five complex, supertranslation-invariant quantities defined on null infinity $\scri^+$ and although put forward in the 60's, they have never been computed in a non-stationary setting. We accomplish this through a numerical implementation of Friedrich's generalized conformal field equations whose semi-global evolution yields direct access to $\scri^+$. Generalizations of the NP constants' integral expressions are made to allow their computation in a more general gauge that better suits the output of a numerical evolution. Canonical methods of fixing inherent degrees of freedom in their definitions are discussed. The NP constants are then computed for a variety of different ingoing wave profiles in axisymmetry, and then with no symmetry assumptions in 3+1 for which all five are non-zero.

\end{abstract}
% \keywords{Newman Penrose constants, null infinity, conformal field equations}
\submitto{\CQG}
\maketitle

\section{Introduction}

Gravitational waves are a robust prediction of general relativity. The existence
of wave solutions to the field equations has been known since the early days of
the theory, but there was doubt that wave-like behaviour occurred generically
outside of overly-symmetric exact solutions \cite{Rosen:1937}. The way we
characterise radiation today emerged out of the work of Bondi
\cite{Bondi:1962}, Sachs \cite{Sachs:1962a}, Newman and
Penrose \cite{Penrose:1963,Newman:1962,Newman:1962b}.
Penrose's procedure of conformal compactification \cite{Penrose:1965}
succinctly encodes the asymptotic fall-off conditions hard-coded by Bondi and
Sachs. The conformal boundary, $\scri$, emerges as the natural place to define
gravitational radiation.

This picture of gravitational radiation owes a great debt to Maxwell's
electromagnetism. The isolation of radiative degrees of freedom by Bondi and
Sachs echoes an earlier analysis of electromagnetic radiation via the
Li\'enard-Wiechert potential and Penrose's conformal compactification is
premised on the conformal-invariance of zero-rest-mass fields, a fact which was
shown for a Maxwell field much earlier
\cite{cunningham1910principle,bateman1910transformation}.

This paper is focused on the explicit calculation of another such import from
Maxwell electromagnetism, the Kirchhoff integral formula. Generalised to fields
of spin-$s$ in Minkowski space, it is called the generalised
Kirchhoff-d'Adh\'emar formula and relates the value of the field at a point to
an integral over an arbitrary smooth \textit{cut} of its (past) light cone. Formally applied to
the spin-2 Weyl spinor on the conformal boundary $\scri^{+}$ of an asymptotically flat
spacetime, the Kirchhoff-d'Adh\'emar formula yields a set of five
complex supertranslation invariant quantities on $\scri^+$ which are ostensibly
the components of the Weyl spinor at timelike infinity. These are the
NP constants \cite{Newman:1968}, whose physical
interpretation has proved elusive for over half a century.

Nevertheless, much has been said about these constants in the intervening years.
In their original paper, Newman and Penrose make the argument that the existence
of the constants has non-trivial physical significance
\cite{Newman:1968,Penrose:1986}. It has since been shown that the
vanishing of the NP constants distinguishes between fundamentally
different late-time behaviour of self-gravitating waves
\cite{Gomez:1994a}. They play an important role in the early
radiative properties of isolated systems close to spatial infinity
\cite{Kroon:2003}. Another line of analysis has found that the NP
constants appear as subleading BMS charges \cite{Godazgar:2019}. However, as far
as we are aware, the NP constants have never been explicitly
computed in a general space-time. 

Explicit numerical computation of quantities at the conformal boundary without
the use of limiting procedures can be done by employing conformal
compactification in a numerical scheme. Friedrich's conformal field equations
regularly extend the Einstein equations to include the conformal boundary
\cite{Friedrich:1981,Friedrich:1981a,Friedrich:1995}.
Recently, an initial boundary value problem (IBVP) framework for the generalised
conformal field equations (GCFE) was presented \cite{Beyer:2017}. This
framework puts $\scri^+$ within the computational domain, allowing for the
non-linear perturbation of black hole space-times.

Because the computational domain includes at least a portion of the future null
boundary, quantities defined there can be computed with local differential
geometrical methods. Most asymptotic quantities, including the NP
constants, are defined on a 2-dimensional \textit{cut} of $\scri^+$, therefore
one can see how a quantity evolves along the set of successive cuts. However, in
the literature, these quantities are often defined in terms of a very specific
set of coordinates, frame, and conformal factor. These choices are usually
incompatible with the requirements of the numerical scheme. Therefore, to
compute a quantity at null infinity with this scheme, it must be written in a
conformally invariant way.

The aim of this paper is to use the numerical framework provided by the IBVP
formulation of the GCFE to compute for the first time, the NP
constants explicitly on $\scri^+$. The case considered is the non-linear
perturbation of a Schwarzschild space-time by gravitational waves. The reader is
referred to \cite{Beyer:2017} for details of the numerical scheme and
checks of correctness such as constraint convergence tests.

The layout of the paper is as follows: Section \ref{sec:gcfe} summarizes the
IBVP framework for the GCFE. Section \ref{sec:npc} presents the NP
constants and proves their supertranslation-invariance in the general form
required for their computation. Section \ref{sec:calc} presents the details of
aligning the frame of the GCFE with the frame in which the NP
constants are defined and discusses the details of their calculation. Section
\ref{sec:results} presents numerical checks of correctness and results for a
range of initial wave profiles and Section \ref{sec:discussion} concludes with a
brief discussion. We follow conventions of Penrose and Rindler
\cite{Penrose:1986} throughout.

\section{Overview of the GCFE and its numerical IBVP implementation}
\label{sec:gcfe}

We implement Friedrich's generalized conformal field equations analogously to
previous papers in this series \cite{Frauendiener:2021a} and here just give a
brief overview.

The conformal field equations are a regular extension of the Einstein equations defined on a physical space-time to another conformally related Lorentzian manifold, related by a conformal factor $\Theta$, where the points at 'infinity' of the physical space-time are given by $\Theta=0$. Imposing the conformal Gau\ss\;gauge on the GCFE~\cite{Friedrich:1995} yields a system of evolution equations of which most are ordinary differential equations except those governing the components of the 
gravitational tensor which form a symmetric hyperbolic system.
These evolution equations are complemented by a set of constraint equations which are preserved by the evolution. The
associated IBVP is completed by constraint
preserving boundary conditions \cite{Beyer:2017} which are used to
generate fully non-linear gravitational dynamics.

The Schwarzschild space-time of mass $m$ written in isotropic coordinates is
again used as the initial space-time. The specific choice of conformal
Gau\ss\;gauge given by Friedrich \cite{Friedrich:2003} is used by which regular coordinates, frame and conformal factor up to and beyond null infinity can be defined.

Our numerical implementation is capable of general $3+1$ dimensional simulations and we use this capability to generate a complete set of non-trivial NP constants going beyond the axisymmetric case.

For all simulations presented here (excluding convergence tests) we use
coordinates $\{t,r,\theta,\phi\}$\footnote{The axisymmetric numerical
implementation is analogous to the proceeding outline but without the
$\phi$-direction and with optimized spin-weighted spherical harmonic
transformations and corresponding $\eth$-calculus calculations.}. The spatial
coordinates are discretized into equidistant points in the intervals
$r\in[m/2, m/2 + 2m]$, $\theta\in[0,\pi)$ and $\phi\in[0,2\pi]$ with 401, 33 and
64 points respectively. The temporal discretization is also equidistant in this
study with timestep given by $dt = dr/2$ giving a Courant-Friedrichs-Lewy number
of 0.5. The MPI-parallelized Python package COFFEE \cite{Doulis:2019}
contains all the necessary numerical methods to evolve this initial boundary
value problem. The standard explicit Runge-Kutta fourth order method is used to march in
time, Strand's fourth order summation-by-parts finite difference operator (third
order on the boundary) \cite{Strand:1994} is used to approximate radial
derivatives and the simultaneous-approximation-method \cite{Carpenter:1994}
is used to stably impose maximally dissipative boundary conditions. Finally, we
use spin-weighted spherical harmonics to allow for fast and accurate angular
derivatives through a pseudo-spectral implementation of Penrose's
$\eth$-calculus \cite{Beyer:2016}. Regridding is also performed, whereby
regions outside of future null infinity are chopped away from the computational
domain to maintain a stable evolution. This is also performed inside the black
hole to avoid the singularity.

\section{Newman-Penrose constants} \label{sec:npc}

The Kirchhoff-d'Adh\'emar construction in Minkowski space expresses a solution
to the zero-rest-mass field equations at a point $P$ as an integral of an
arbitrary smooth \textit{cut} of the (past) light cone of $P$ \cite{Penrose:1984a}. It is a
conformally invariant construction and so we may apply it specifically to relate
a zero-rest-mass field $\phi_{AB\ldots L}$ at the future (past) timelike infinity of a
conformally rescaled spacetime to a \textit{cut} of its light cone, future (past)
null infinity.  Because the construction is invariant with respect to the cut of
the light cone on which it is evaluated, it gives a set of
supertranslation invariant constants corresponding to the $2s+1$ components of
the zero-rest-mass field at timelike infinity. In a spacetime with matter,
timelike infinity becomes a singular point of the conformally rescaled manifold,
but the integrand of the Kirchoff-d'Adh\'emar construction, being evaluated on null infinity, remains regular and
so applied to the spin-2 Weyl spinor, we are left with a set of five complex
supertranslation invariant quantities defined on any asymptotically flat spacetime.
These are absolutely conserved in the sense that they remain constant even with 
non-vanishing news.

The physical set-up of the Kirchoff-d'Adh\'emar construction is as follows.
Consider a point $P$ with a (past or future) light cone $\mathscr{I}$ and an
arbitrary smooth 3-dimensional null hypersurface $\mathscr{N}$ intersecting
$\mathscr{I}$. The intersection has spherical topology and is labeled $\mathscr{C}$.
The Kirchhoff-d'Adh\'emar construction will take the form of an integral
evaluated on $\mathscr{C}$ whose value is independent of the intersecting null
hypersurface $\mathscr{N}$ and  thus of the specific intersection $\mathscr{C}$. In
fact, any smooth cut of the light cone $\mathscr{I}$ can be said to have come
about by the intersection of $\mathscr{I}$ with some null hypersurface
$\mathscr{N}$.

The method of proof consists in showing that the Kirchhoff-d'Adh\'emar integral
evaluated on two arbitrary cuts of $\mathscr{I}$ denoted $\mathscr{C}$ and
$\mathscr{C}'$ gives the same result by treating $\mathscr{C} - \mathscr{C}'$ as
the oriented boundary of a 3-dimensional section of the light cone
$\mathscr{I}$. The generalised Stokes' theorem can be used to relate
cut invariance to the vanishing of a related integral on the region between the
cuts. This is explained in detail in \cite{Penrose:1986,Newman:1968}

At each point of an intersection $\mathscr{C}$ there are two distinguished null directions,
one along the generators of the light cone $\mathscr{I}$ and the other along the intersecting null
hypersurface so it is advantageous to use the GHP formalism~\cite{Geroch:1973b}. We can choose
a spin-frame such that $o^A$ points along the intersecting hypersurface,
and $\iota^A$ along $\scri$, normalised so that $o_A \iota^A = 1$.

Formally, the Kirchhoff-d'Adh\'emar formula reads 
\begin{equation}
    \phi[U]\bigm{|}_P = \int_\mathscr{C} U \thorn_{c}\phi\ \mathbf{d^2 \mathcal{C}}
    \label{eq:kda}
\end{equation}
where $\phi := \phi_{AB\ldots L}o^Ao^B\ldots o^L$, $U$ is a weighted scalar satisfying
\begin{equation}
\label{eq:Ucondn}
(i) \; \bar{\eth}_c U = 0, \quad \text{ and }\quad (ii) \; \thorn_c' U =0.
\end{equation}
and the derivative operators with a subscript 'c' are conformally weighted operators 
of the cGHP formalism as introduced in~\cite{Frauendiener:2022}.

In Minkowski spacetime, in a gauge where each cut $\mathscr{C}$ is represented as a unit 2-sphere, $U$ will be a component of the spinor $\iota_A\iota_B\ldots\iota_L$, i.e., one of the $2s+1$ spin-weighted spherical harmonics ${}_{-s}Y_{sm}$ with spin-weight $-s$. In this gauge, these are constant, thus trivially propagating along the light cone.

In a curved spacetime, $U$ is a generalisation of these spin-weighted spherical harmonics to a topological but not necessarily metric sphere. In this general case, there are still $2s+1$ independent solutions of (\ref{eq:Ucondn}(i)) for $U$.

\section{Calculating the NP constants} \label{sec:calc}

In the remainder of this work we focus on the gravitational NP constants, i.e., with $\phi_{ABCD} = \psi_{ABCD}$, the conformally rescaled Weyl spinor. Many system variables of the GCFE are components of spinors with respect to a
certain spin-frame. In general, this spin-frame, and its associated null tetrad,does not agree with the frame adapted to $\scri$ that was used in the definition of the
NP constants but we can use null rotations to transform between the
GCFE null frame and the frame adapted to $\scri$ herein referred to as a
Bondi frame\footnote{This is not strictly correct, since we are referring here only to a single cut, whereas the standard usage of the term Bondi frame refers to an entire system of cuts parametrised by the retarded Bondi time.}.

A null rotation mixes one component of a spin-frame into
another. For example, a null rotation of $o^A$ around $\iota^A$ is given by
\begin{align}
  \label{eq:nullrot}
  & o^{A} \to o^{A}+ Y\iota^{A},\quad \iota^A \to \iota^A
\end{align}
thus keeping $\iota^A$ fixed, where $Y$ is a function of the spacetime coordinates. If we denote the
Bondi spin-frame by $O^{A}$ and $I^{A}$, the GCFE frame by $o^{A}$ and
$\iota^{A}$, and the corresponding Bondi and GCFE null-tetrad vectors by capital
and lowercase letters respectively, then we may transform between frames by two null successive rotations (first fixing $o^A$ and then the new $\iota^A$) which have the combined form
\begin{align}
  \label{eq:doublenullrot}
  & O^{A} = o^{A}+ Y(\iota^{A}+Xo^{A}), & I^{A} = \iota^{A}+ Xo^{A}.
\end{align}
The null rotation functions $X$ and $Y$ are determined by the conditions
\begin{align}
  \label{eq:nullrotcondn}
  & \nabla_{a}\Theta = -AN_{a}, & M^{a}\nabla_{a}t = 0,
\end{align}
where the scaling $A$ is fixed given the conformal factor $\Theta$ and the above expression of the adapted spin-frame. These conditions impose that $N^{a}$ points along the null generators of $\scri$ and that the complex vector $M^{a}$ lies within the $t=\text{const.}$ cuts of $\scri$. Appropriately fixing the freedom in how the frame propagates along the timelike conformal geodesics of the conformal Gau\ss\ gauge allows one to satisfy the second condition automatically, yielding $X=0$. The first condition gives us the
value of the null rotation function $Y$ on $\scri^+$. 

The transformation between the GCFE frame and the Bondi frame is then known and so we may write components with respect to the Bondi frame in terms of components with respect to the GCFE frame which are known numerically. As a simple example, the third component of the gravitational spinor is written in the Bondi frame as
\begin{equation*}
  \psi_{ABCD}O^{A}I^{B}I^{C}I^{D} = \psi_{ABCD}(o^{A}+Y\iota^{A})\iota^{B}\iota^{C}\iota^{D} = \psi_{3} + Y\psi_{4}.
\end{equation*}
Both the GCFE and the NP constants are defined with respect to a spin and boost
covariant formalism and so a properly weighted expression with respect to one
frame  results in a properly weighted expression with respect to another.

The same process is used to compute the area-form in terms of numerically
available quantities.

\subsection{Fixing the behaviour of the frame off $\scri^{+}$}

With the above two null rotations, the frame on $\scri^{+}$ is fixed, but we
also have some freedom to choose how our frame changes as we move away
from $\scri^{+}$. The presentation of $\scri^{+}$ in the proof of
supertranslation invariance makes use of this and takes $\kappa = 0$ since the intersecting null hypersurface is foliated by a null geodetic congruence~\cite{Penrose:1986}.

The Bondi frame so far is only fixed on $\scri^+$ and in order to achieve $\kappa=0$ we need to enforce that $Do^A \propto o^A$ which means that we need to determine the null rotation function $Y$ away from $\scri^+$. Suppose, we have fixed $Y$ on $\scri^+$ with the above procedure, then we can get the required result with a third null rotation that becomes the identity on $\scri^+$
\begin{equation}
  \label{eq:nullrot3}
o^{A} \to \hat{o}^A = o^{A}+Z\iota^{A}, \quad \iota^{A} \to \hat{\iota}^A = \iota^{A}, \quad \text{where } Z = \mathcal{O}(\Theta),
\end{equation}
recalling the conformal factor $\Theta$. Under this transformation,
\begin{equation}
  \label{eq:kappatrans}
  \hat{\kappa} = \kappa - DZ ,
\end{equation}
where $D:=L^a\nabla_a$ and choosing $Z$ so that $\kappa = DZ$ on $\scri^{+}$ we obtain $\hat\kappa = 0$ there. 

Although the transformation becomes the identity on $\scri^+$, we must worry
about derivatives of the frame. In the Kirchhoff-d'Adh\'emar integral, we have a derivative of the form $D\phi$ where $\phi = \phi_{AB..L}o^{A}o^{B}..o^{L}$ ($2s$ indices). Under this null rotation,
\begin{align}
  \hat{D}\hat{\phi} \Bigm|_{\scri^+} &= D\hat{\phi} \\
                                             &= D(\phi_{A..L}(o^{A}+Z\iota^{A})..(o^{L}+Z\iota^{L})) \\
  &= D\phi + 2s\kappa\phi_{1}
\end{align}
since the derivative of any term containing powers of $Z$ higher than one will
vanish on~$\scri^{+}$.

\subsection{Computing $U$}

In the NP constant integrand \eqref{eq:kda}, the active component
would appear to be the term $\thorn_{c}\phi$ since this brings information of
the arrival of the field $\phi$ at $\scri^+$, while the quantity $U$ appears to
be somewhat inert, being used to project out certain pieces of information from the integrand. Before the jump was made to curved spacetime, the
Kirchhoff-d'Adh\'emar integral could represent the value of the field $\phi$ at
timelike infinity. In this case, the quantity $U$ was replaced by components of
$\iota_{A}\iota_B...\iota_L$ which are spin-weighted spherical harmonics in an
appropriate frame. When curved spacetime is introduced, the $U$ takes the role
of these components and so different choices of $U$ which satisfy the underlying
equations~\eqref{eq:Ucondn}, represent what would have been components of
$\phi$ at timelike infinity. The job of $U$ is to lend its spin, boost, and
conformal weight to the expression, and so provide alignment between cuts of
$\scri^+$ allowing for comparison from cut to cut.

To compute $U$ we must solve the ``constraint equation'' $\bar\eth_cU = 0$ on a
cut\footnote{This term is justified since, by considering the commutator $[\thorn_c',\bar\eth_c]$ one can show that any $U$ satisfying the evolution equation $\thorn_c'U=0$ will satisfy $\bar\eth_cU=0$ on every cut if it satisfies this equation on a single cut. In this sense, the constraint is propagated by the evolution.} and evolve it along the null generators of $\scri^{+}$ with the evolution
equation $\thorn_{c}'U = 0$. Following methods from our earlier paper \cite{Frauendiener:2021a}, we can expand these operators written in the Bondi frame in terms of the numerically implemented, standard operators $\tilde{\eth},\tilde{\eth}'$, to obtain
\begin{equation}
  \label{eq:numericalUconstraint}
 A\tilde{\eth}U + B\tilde{\eth}'U + CU = 0.
\end{equation}
Expanding the known coefficients $A$, $B$, and $C$, and
the unknown function $U$ in
terms of spin-weighted spherical harmonics, and using the well-known relationship for
products of spin-weighted spherical harmonics in terms of Clebsch-Gordon
coefficients (see~\cite{Penrose:1984a}) results in a system of homogeneous linear equations for the spectral
coefficients of the function $U$. There are five linearly independent solutions which span the
solution space to the constraint equation for $U$.

The evolution equation can similarly be written in terms of numerically
available quantities as,
\begin{equation}
  \label{eq:num_ev}
  A \partial_t U + B \tilde\eth U + C \tilde\eth' U + DU = 0,
\end{equation}
and may be evolved along $\scri^{+}$ by the method of lines given an initial solution to
the constraint equation. An adaptive fourth-order Runge-Kutta method is used since the numerical output is not linearly spaced in $t$.

\subsection{Fixing a basis in the solution space $\mathscr{U}$}\label{sec:fixing-basis-solut}

It is clear that solution $U$ of (\ref{eq:Ucondn}(ii)) leads to another solution $\alpha U$ provided that $\alpha$ is a complex constant on the cut $\mathscr{C}$.  More generally, any complex linear combination of solutions will also be a solution. Thus, changing the basis of the solution space  $\mathscr{U}$ of (\ref{eq:Ucondn}(ii)) will also change the individual values of the five NP constants. Therefore, these do not, by themselves, carry independent physical information. Only the combination of the values of the integrals together with the knowledge of the basis of $\mathscr{U}$ carries the full information.

This means that in order to compare NP constants across different spacetimes we need to make sure that we specify ``the same basis'' for the solution space for each spacetime. There are several ways to do this: two complicated ones which are also physically relevant, and a third easier one which not as physically meaningful but much more pragmatic.

The first idea that comes to mind is to first conformally rescale the metric on the cut to make it into a unit-sphere, and, in a second step, to change the coordinate system by a Möbius transformation so that it becomes a standard polar coordinate system on the unit-sphere. In this situation, the solutions of (\ref{eq:Ucondn}(ii)) are the standard spin-weighted spherical harmonics $Y_m:={}_{-2}Y_{2m}$ with $-2 \le m \le 2$.

While the first step is rather straightforward, the second step leads to a Poisson-type equation on the sphere with a $\delta$-like source term which is difficult (but not impossible) to treat numerically. In addition, after the coordinate transformation all quantities must be transformed which may introduce several numerical errors.

The second way to introduce a basis in $\mathscr{U}$ is to make use of the fact that the standard spin $2$ spherical harmonics $Y_m$ form an irreducible representation of the group $SU(2)$. They each are an eigenvector of an infinitesimal generator with different eigenvalue, and they are obtained one from the other by the action of two ladder operators (very much akin to the angular momentum algebra of quantum mechanics). Fixing one of them as being annihilated by one of the ladder operators, one can generate the others by successive application of the other ladder operator. This fixes the complete basis in terms of the first vector and leaves the freedom of scaling with one complex number. This can be almost fixed by normalising the vectors with respect to an appropriate Hermitian product, leaving the remaining freedom of a single phase.

In principle, this program could be carried out but it is very cumbersome. First, one needs to find the infinitesimal generators of the group action. This leads to a series of elliptic equations to be solved on the sphere. Next, one needs to use these generators to determining the function that is killed by one of the ladder operators, which is again an elliptic equation on the sphere, and then generate the other functions by successive application of the other ladder operator. As an alternative, one could solve the eigenvalue problem for the third operator. Obviously, this procedure is numerically quite involved and prone to inaccuracies due to successive numerical differentiation.

For this reason, we use a third method to fix a ``universal'' basis of $\mathscr{U}$, and we use the ``universal structure'' that is available to us, namely our numerical setup which is the same for every spacetime that we compute. Recall that our method is based on concentric round spheres and that every function we compute can be expanded as a linear combination of spin-weighted spherical harmonics defined with respect to the numerical round spheres. Therefore, we proceed as follows: first, on an initial cut, we compute five linearly independent solutions $(u_k)_{k=-2:2}$ of (\ref{eq:Ucondn}(ii)). These have the form
\[
  u_k = \sum_{m=-2}^2 c^m{}_k Y_m + Z_{l>2}, \quad k=-2:2
\]
where $Z_{l>2}$ stands for terms with higher values of $l$. Then a straightforward linear combination of these solutions leads to the universal basis $U_k$ which is defined by
\[
  U_k = Y_k + Z_{l>2}  
\]
where $Z_{l>2}$ again stands for higher $l$ terms. We can interpret this basis as being the deformation of the standard basis provided by the $Y_m$ due to the impact of the incoming gravitational wave. If there was no gravitational wave, then the cut would be spherically symmetric and the $y_k$ would agree with the standard basis. The basis, thus defined, is then propagated along $\scri^+$ using the evolution equation (\ref{eq:Ucondn}(i)). In this process, the form of the $y_k$ will change. This process of fixing a ``universal basis'' of $\mathscr{U}$ leaves no further freedom (except, of course, for the free phase inherent in the definition of the spin-weighted spherical harmonics).

\subsection{Integrating the Newman-Penrose integrand}

At this stage, all elements of the Newman-Penrose integral~\eqref{eq:kda} have expressions in terms of known quantities. Integration can be performed against the basis $U_k$ as defined in~\ref{sec:fixing-basis-solut} by simply computing the $s=l=m=0$ spectral coefficient of the complete integrand and dividing by $2\pi$. The theory shows these five complex numbers, obtained on a cut, come out the same independently of which cut was chosen for their evaluation. In the next section we present numerical results that showcase these properties.

\section{Numerical Results} \label{sec:results}

Using the above procedure, the NP constants were computed with data on $\mathscr{I}^+$ from a numerically evolved spacetime modelling the non-linear perturbation of a Schwarzschild black hole by an incoming gravitational wave pulse.  Because the NP constants are evaluated  on each cut of $\mathscr{I}^+$ defined as the intersection with a $t=\const$ hypersurface, we obtain five complex numbers for every $t$.

The initial mass of the black hole is $m=0.5$ for all simulations considered.

\subsection{Ingoing wave}

The ingoing pulse is defined as the choice of free data $q_0$ of the lightlike, ingoing characteristic variable on the outer boundary. This is chosen to be a linear combination of the spin-weighted spherical harmonics ${}_2Y_{2m}$ for $m=0,1,2$ and with amplitudes $a$, $b$ and $c$ respectively. The choices of these amplitudes will vary in the upcoming sections. This gives the wave profile on the outer boundary
\begin{equation}
	q_0(t,\theta,\phi) =
	\begin{cases} 
	    [4a\sqrt{\frac{2\pi}{15}}\;{}_2Y_{20}
      - 2b\sqrt{\frac{5}{\pi}}\;{}_2Y_{21}
      + 2c\sqrt{\frac{\pi}{5}}\;{}_2Y_{22}]
      \sin^8(8{\pi t }) & t \leq\frac18 \nonumber \\
	    0 & t >\frac18
	\end{cases}.\label{eq:WaveProfile}
\end{equation}

\subsection{Checks of correctness}
We have demonstrated in several papers~\cite{Beyer:2017,Frauendiener:2021a} that the solutions computed by the GCFE system converge at the correct order for the $2+1$ axisymmetric case. Here we show that this is also true in the general case of $3+1$ dimensions. We also show that the NP constants converge to constant values on $\scri^+$.
\begin{figure}[h]
  \centering
  \begin{subfigure}[t]{0.3\textwidth}
    \centering
    \includegraphics[height=1.2in]{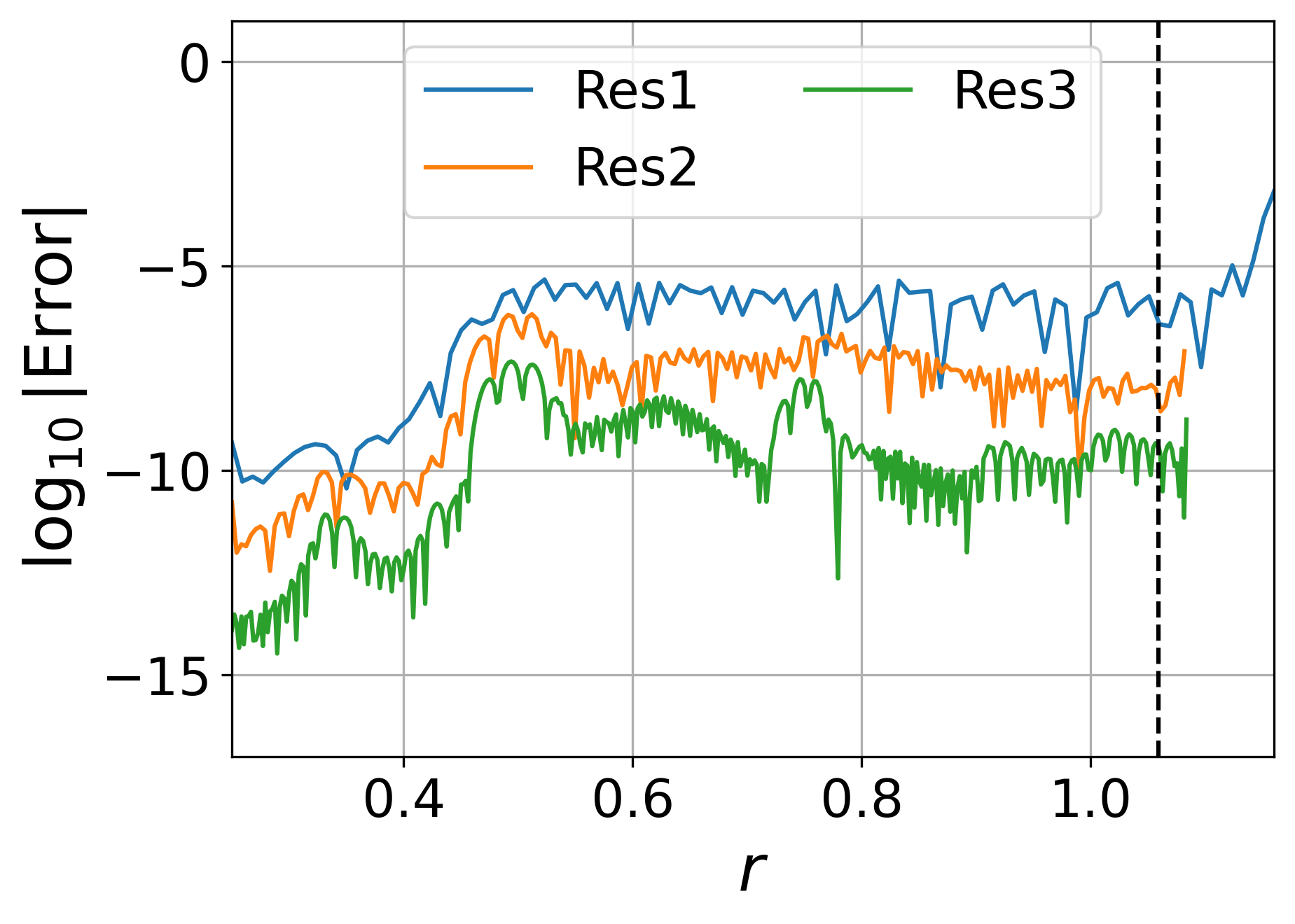}
    \caption{$\theta=\pi/2$ and $\phi=\pi$.}
  \end{subfigure}
  ~
  \begin{subfigure}[t]{0.3\textwidth}
    \centering
    \includegraphics[height=1.2in]{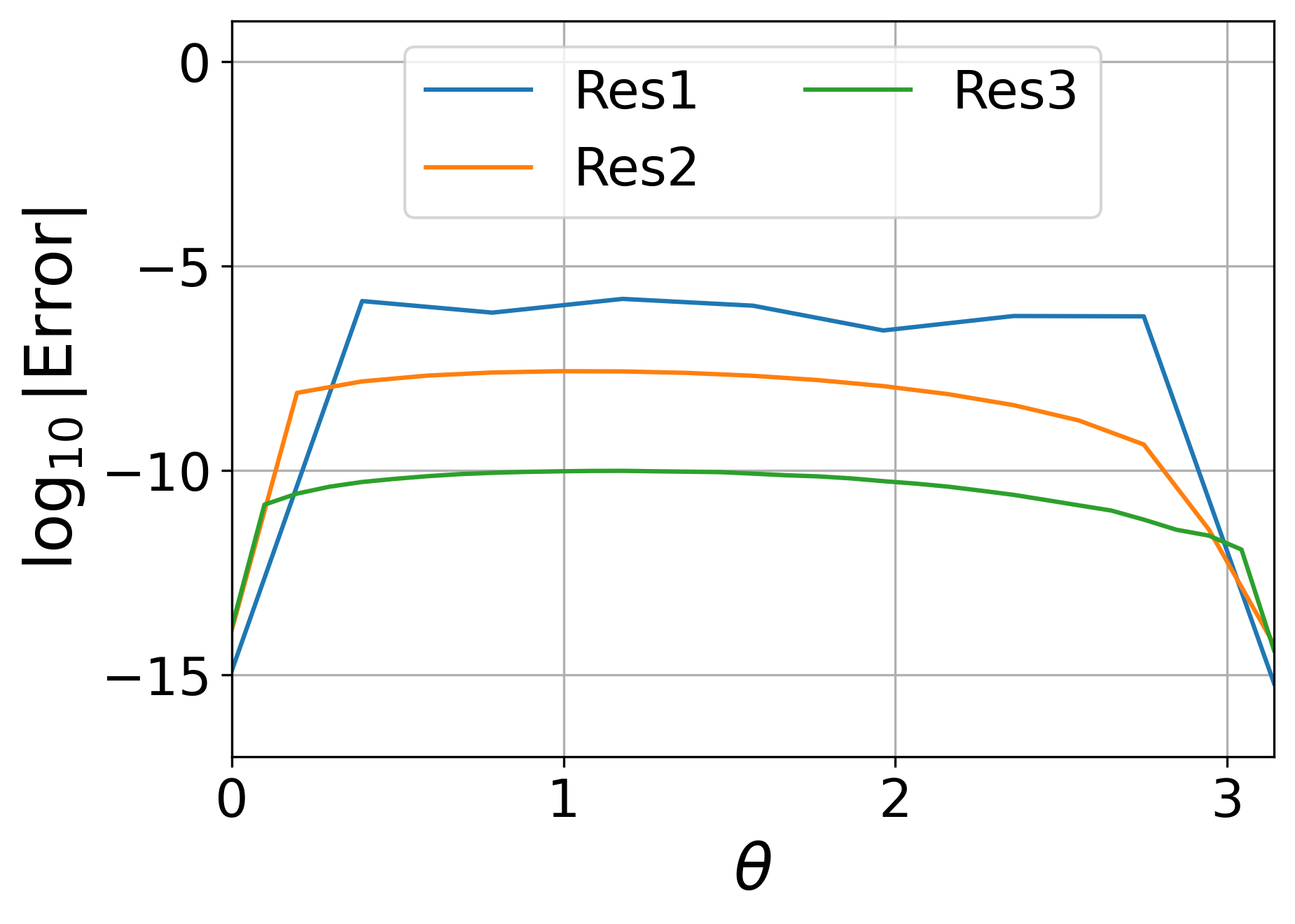}
    \caption{$r=0.978$ and $\phi=\pi$.}
  \end{subfigure}
  ~
  \begin{subfigure}[t]{0.3\textwidth}
      \centering
      \includegraphics[height=1.2in]{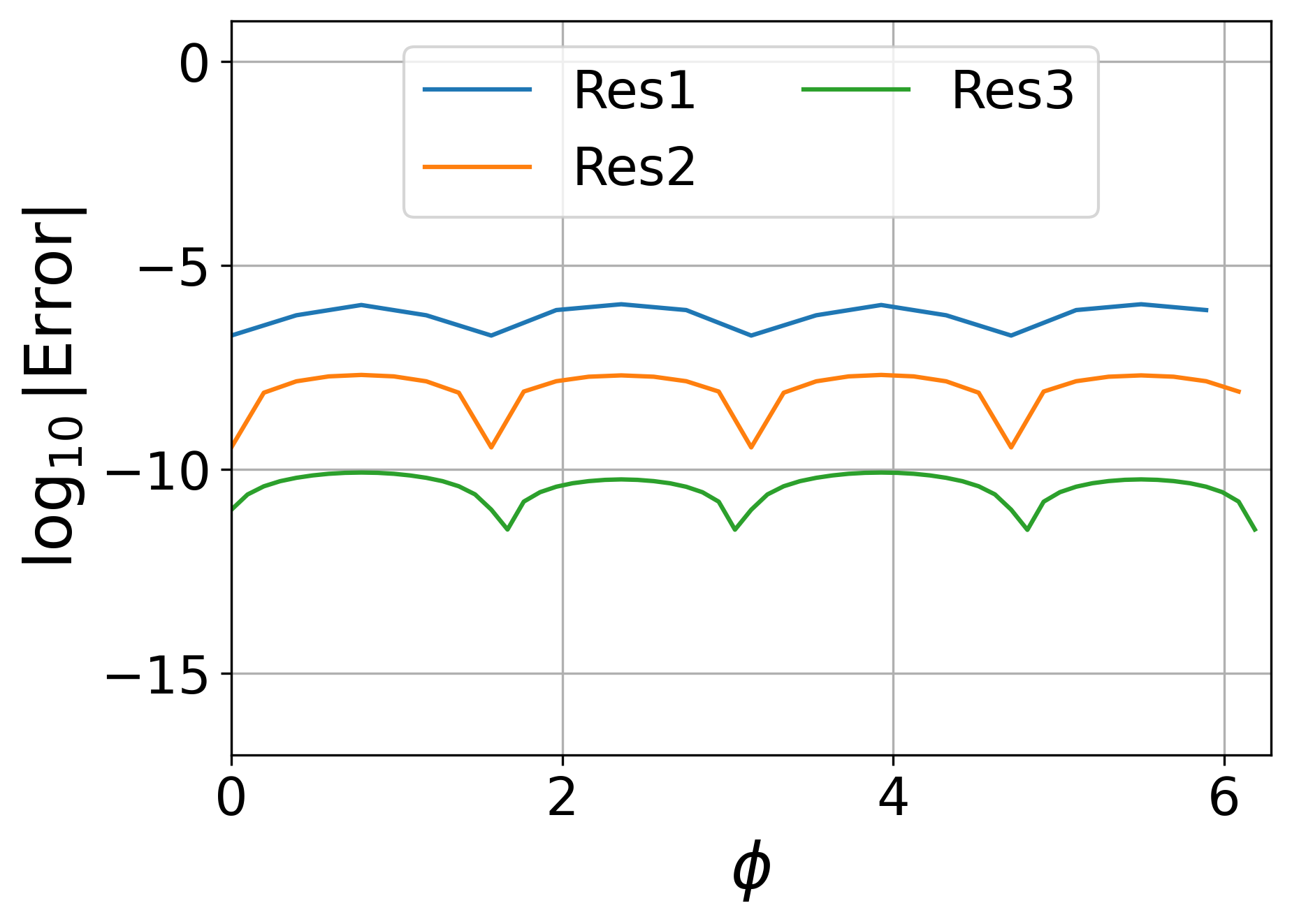}
      \caption{$r=0.978$ and $\theta=\pi/2$.}
  \end{subfigure}
  \caption{The imaginary part of a constraint equation from the Bianchi identity when a single spatial coordinate is fixed at $t=1$ for $r,\theta,\phi$ resolutions of $\{101, 9, 16\},\,\{201, 17, 32\}$ and $\{401, 33, 64\}$ (denoted by Res1, Res2 and Res3 respectively). The dashed vertical line represents $\scri^+$ in the radial plot. The curves from top to bottom correspond to increasing resolution.}\label{fig:4DConvTest}
\end{figure}

We use an ingoing wave profile proportional to ${}_2Y_{22}$ with $a=b=0$, $c=\ii$ to allow excitation in the $\phi$-direction. Fig.~\ref{fig:4DConvTest} demonstrates convergence in all spatial directions at $t=1$.

\begin{figure}[h]
\centering
\includegraphics[width=0.8\textwidth]{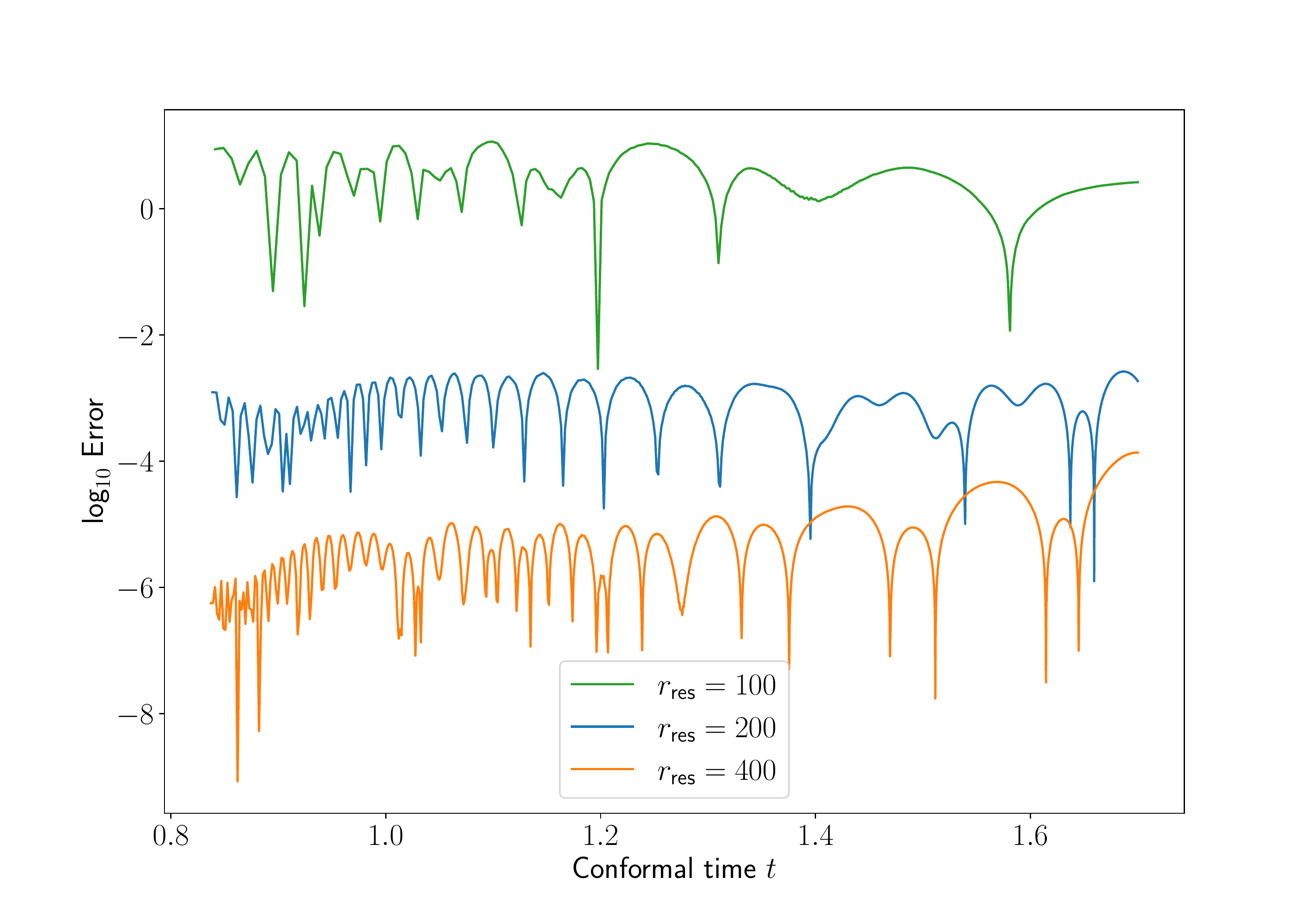}
\caption{Convergence of the $\text{log}_{10}$ difference between the magnitude of
  the NP constant at time $t$ and at the initial cut with increasing spatial resolution.}
\label{fig:convergence}
\end{figure}

Focusing now on the NP constants, we demonstrate how they approach constant values along $\scri^+$. Fig. \ref{fig:convergence} shows a convergence test of the discrepancy from constancy of the single non-vanishing NP constant in axisymmetry, choosing amplitudes $a=1$, $b=c=0$, along $\scri^+$ as the spatial resolution is increased. The resolution in $r$ (number of intervals) is denoted $r_{\text{res}}$ and corresponds to an equivalently scaled resolution $\theta_{\text{res}}$ along the $\theta$-direction. The coarsest values are $r_{\text{res}} = 100$ and $\theta_{\text{res}} = 8$, and the resolutions are doubled in successive simulations.

\subsection{Variable ingoing amplitude}

An superficial glance at \eqref{eq:kda} would suggest that due to the linearity of the integrand and the zero-rest-mass field equations with respect to $\phi$, scaling the amplitude of the ingoing wave would just scale the NP constants linearly. However, this turns out not to be the case because $\phi$ satisfies the Bianchi equation into which $\phi$ enters non-linearly through the connection coefficients of the covariant derivative \cite[§5.7]{Penrose:1984a}. Hence, it is interesting to numerically probe the scaling of the NP constants as the ingoing wave profile is scaled.

\begin{figure}[h]
\centering
\includegraphics[width=0.8\textwidth]{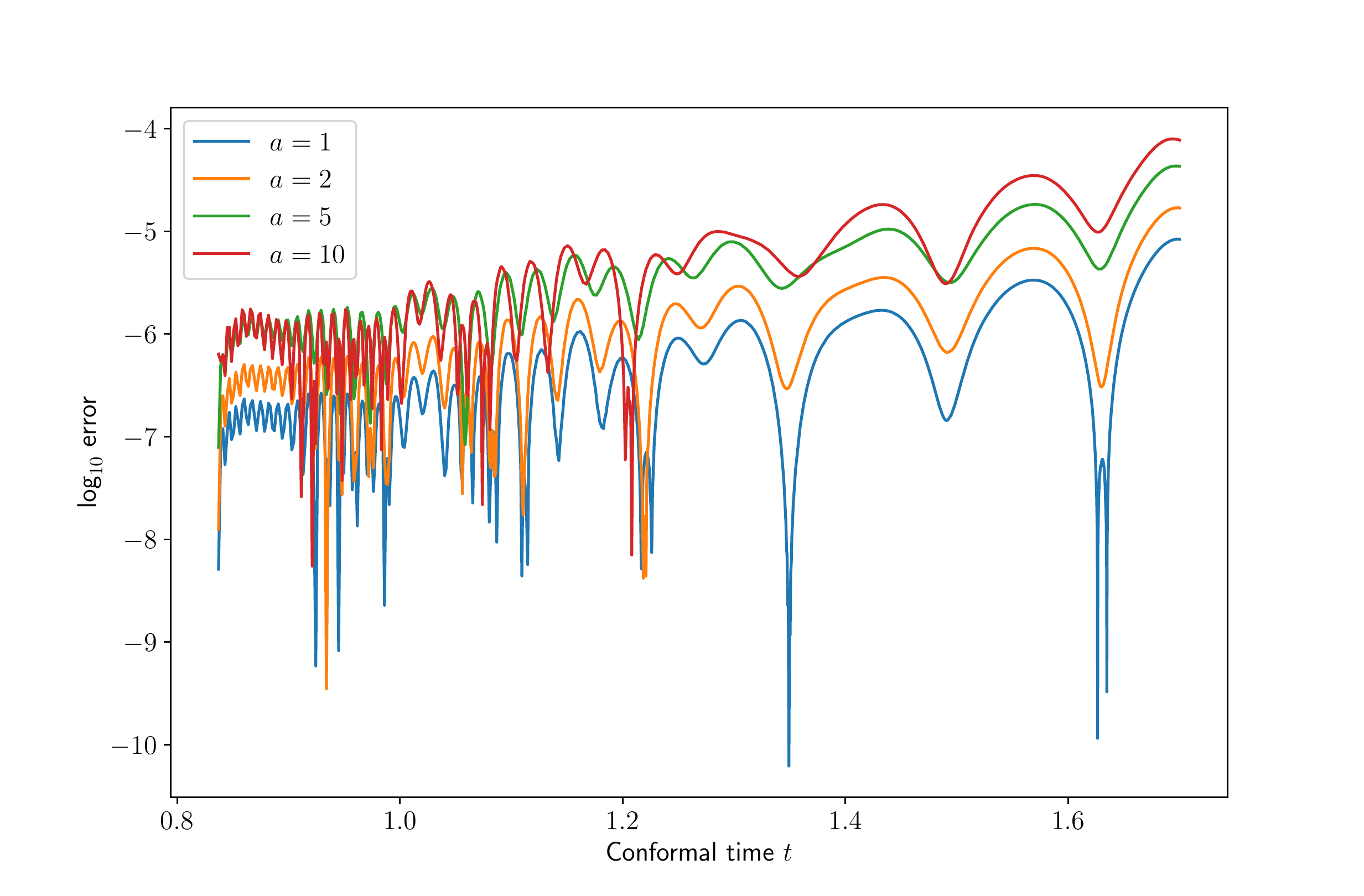}
\caption{The $\text{log}_{10}$ difference of the NP constant from the value on an initial cut as a function of conformal time $t$ for a variable amplitude of the initial wave profile as a measure of deviation from constancy due to error. Cumulative error grows as we integrate along $\scri^{+}$.}
\label{fig:varamp_err}
\end{figure}

We performed four simulations in axisymmetry with the amplitudes $b=c=0$ and $a$ taking the values 1,2,5, and 10. These are evolved up to $t=1.77$ at which point the system variables start to diverge due to the close `conformal' proximity to $i^+$ at $t\approx1.79$. Table \ref{tab:varamp} shows the corresponding single NP constant for each amplitude as well as the relative error from a linear fit through the origin. Fitting the NP constants to the ansatz $\alpha a^{\beta}$ yields $\alpha \approx 0.53865$ and $\beta \approx 0.99803$. Fig. \ref{fig:varamp_err} shows the $\text{log}_{10}$ deviation of the NP constant value from the value on the initial cut for each amplitude as a function of $t$. The deviation from a linear fit is orders of magnitude greater than the error. This is due to the amplitude of the initial wave profile entering into the field equations non-linearly resulting in a non-linear relationship between initial amplitude and Newman-Penrose constant.

\begin{table}[h]
\begin{center}
\begin{tabular}{c|c|c|c|c}
  Amplitude & 1 & 2 & 5 & 10 \\ \hline
NPC & 0.53882 & 1.07638 & 2.68405 & 5.36222 \\
Rel. Err. & 0 & 0.00116 & 0.00373 & 0.00482
\end{tabular}
\end{center}
\caption{The one non-vanishing NP constant for different ingoing wave amplitudes
and the deviation from a linear fit through the origin. This deviation is orders
of magnitude larger than the error for each. This is a result of the amplitude entering
non-linearly into the field equations.}
\label{tab:varamp}
\end{table}

\subsection{Deviation from axisymmetry}

In a general asymptotically flat spacetime there are five complex NP constants corresponding to the five independent solutions to the equations \eqref{eq:Ucondn}. In axisymmetry, these collapse to only one independent solution, when the frame and coordinates also respect the symmetry, because then only the $m=0$ modes of a spherical harmonic expansion remain.

We can investigate the collapse of five NP constants into one by using the initial wave profile given by~\eqref{eq:WaveProfile} and using $a=\ii$, $b=c=\epsilon\,i$, where $\epsilon$ parametrises a deviation from axisymmetry.

\begin{figure}[h]
\centering
\includegraphics[width=0.8\textwidth]{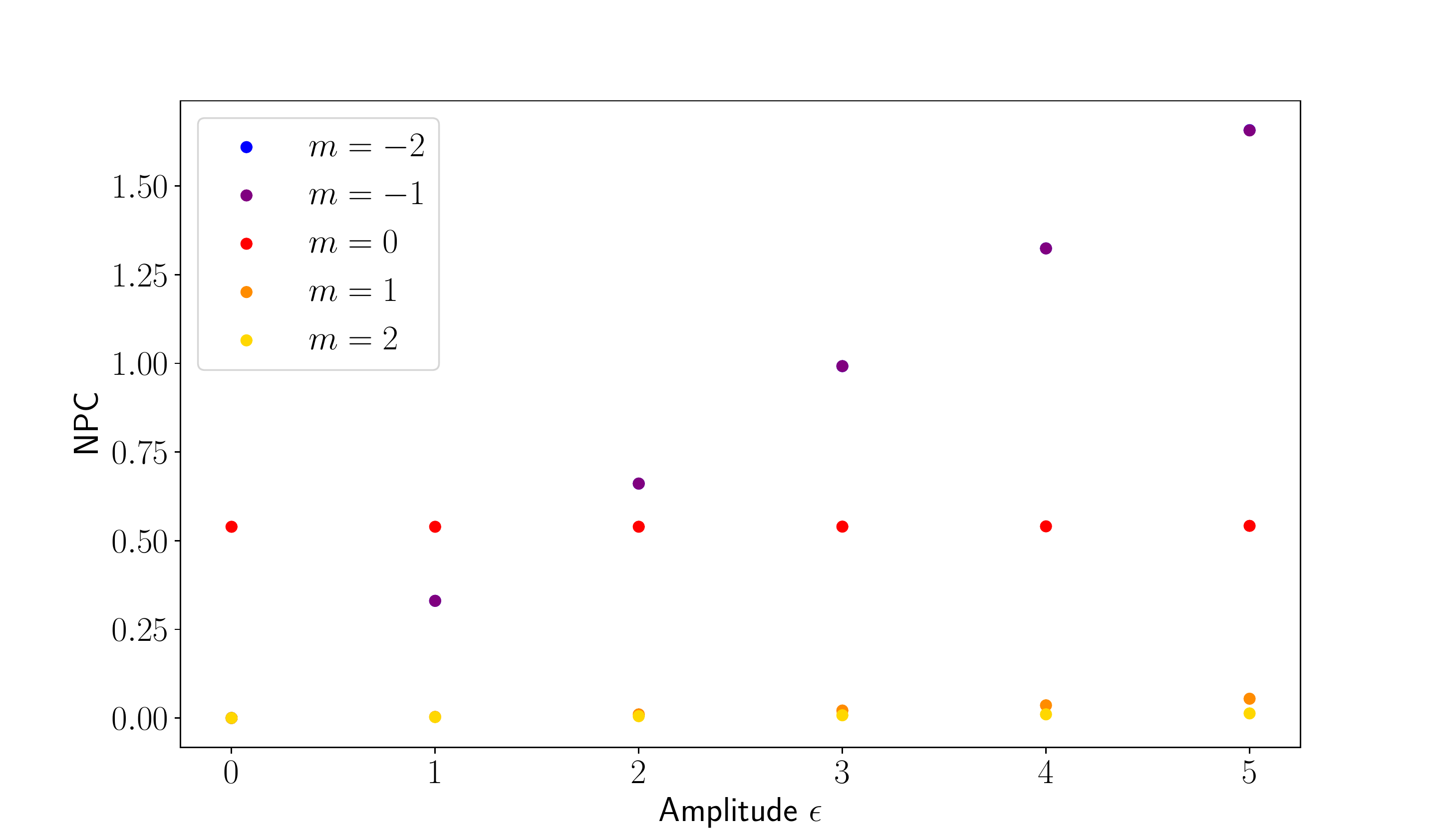}
\caption{The magnitudes of the five complex NP constants as a function of a
parameter $\epsilon$ which breaks axisymmetry in the initial wave profile. For
$\epsilon>0$ all constants are non-zero although most are small.}%
\label{fig:toaxi}
\end{figure}

\begin{figure}[h]
\centering
\includegraphics[width=\textwidth]{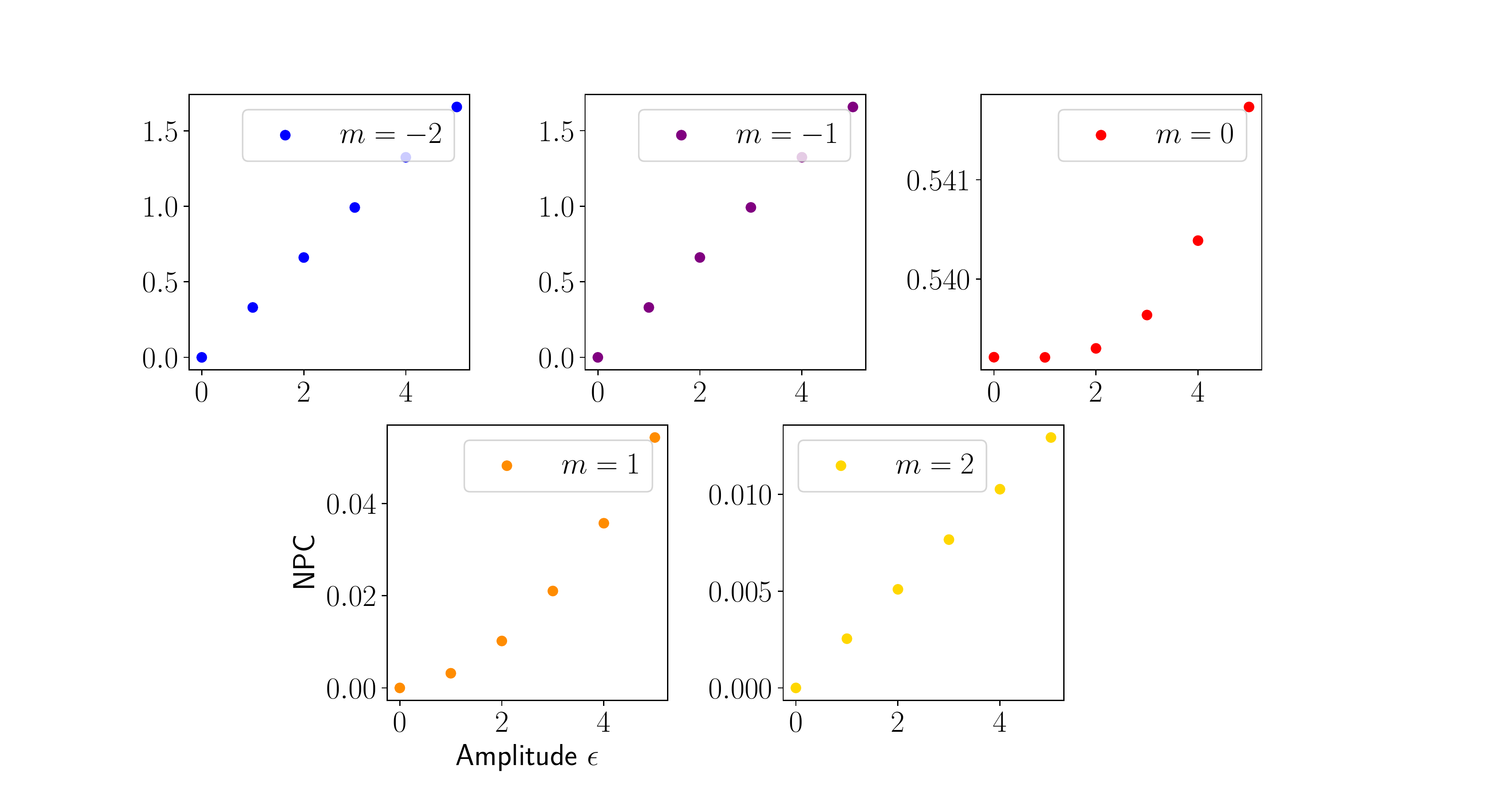}
\caption{The magnitudes of the five complex NP constants as a function of a
parameter $\epsilon$ which breaks axisymmetry in the initial wave profile split
by the value of $m$ of the corresponding $U$. Note that $m$ is a label on spherical harmonics, not a mass.}%
\label{fig:toaxi_bym}
\end{figure}

Six simulations were run for this wave profile for $\epsilon = 0,1,2,3,4,5$. Fig. \ref{fig:toaxi} shows the magnitudes of the corresponding NP constants. Although we do have access to the full ten real degrees of freedom (five complex degrees of freedom) for each simulation, the trends can be seen in the behaviour of the magnitudes. Fig. \ref{fig:toaxi_bym} shows the same quantities but separated by the value of $m$ of the corresponding $U$ so that individual trends can be seen.
We can see that for $\epsilon=0$, there is only one non-trivial constant
corresponding to the axisymmetric $m=0$ mode, but
as non-axisymmetric modes are introduced for $\epsilon>0$, all five constants take
on non-trivial values and grow with $\epsilon$. 

\section{Discussion} \label{sec:discussion}

In this paper, we continue our numerical investigation into the non-linear perturbation of a Schwarzschild black hole using an initial boundary value problem for the general conformal field equations. This novel numerical scheme allows us to include $\scri^+$ within the computational domain and so compute quantities there directly. Thereby, we have computed the NP constants for the first time in a physically realistic spacetime.

The gauge quantities of the system were fixed by numerical needs which implies that we were unable to directly use the very specific set of coordinates, frame, and conformal factor typically used when defining quantities at $\scri^{+}$. To compute physical quantities such as the NP constants with data from the numerical simulation, we need an explicitly conformally invariant expression for the quantity. However, this concept of conformal invariance is a rather special one, and it might be appropriate to highlight it again here. Physical quantities make reference to the \emph{physical metric} $\tilde{g}_{ab}$ of the spacetime. In our context, the physical metric is represented as $\tilde{g}_{ab} = \Omega^{-2} g_{ab}$, i.e., in terms of another metric in the same conformal class and the conformal factor relating the two. By conformal invariance we \emph{do not} mean invariance under $\tilde{g}_{ab}\mapsto \theta^2 \tilde{g}_{ab}$, but rather the invariance under $(g_{ab},\Omega) \mapsto (\theta^2 g_{ab}, \theta \Omega)$, which corresponds to the free choice of the splitting of $\tilde{g}_{ab}$ into a conformal and a scale part.

For example, the recent analysis of the Bondi-Sachs energy-momentum in this framework involved generalising the procedure of constructing a basis of translations with respect to which components of the Bondi-Sachs 4-vector may be taken. The standard procedure of choosing the first four spherical harmonics is certainly not conformally invariant in this sense.  This led to an invariant characterisation of the Lorentzian metric on the space of BMS translations~\cite{Frauendiener:2022}. Of course, the existence of this metric still leaves the freedom of Lorentz transformations for the choice of the basis.

We run into the same problem when defining a basis for the quantity $U$ which,
when integrated against the Newman-Penrose integrand, gives the
linearly independent NP constants. Again, it is the solution space which is defined in a conformally invariant way. But in this case there is no obvious inner product that one could use to select a basis. Even if there was one, the basis would still be defined only up to the appropriate (pseudo-) orthogonal transformations. We circumvent the non-uniqueness of the basis in this case by refering to the universal structure that is imposed on the problem by the numerical setup as explained in Sec.~\ref{sec:fixing-basis-solut}. This seems to be the best way to ensure comparability across the different space-times that we investigate.

\ack
Supported by the Marsden Fund Council from Government funding, managed by Royal Society Te Apārangi.

The authors would like to thank L. Escobar for sharing the general form of his SWSH code.

We wish to acknowledge the use of New Zealand eScience Infrastructure (NeSI) high performance computing facilities, consulting support and/or training services as part of this research. New Zealand's national facilities are provided by NeSI and funded jointly by NeSI's collaborator institutions and through the Ministry of Business, Innovation \& Employment's Research Infrastructure programme. URL \url{https://www.nesi.org.nz}.

% \appendix

% \bibliographystyle{elsarticle-num} 
% \bibliographystyle{cqg} 
\printbibliography
\end{document}